# Record High Magnetic Anisotropy in Chemically Engineered Iridium Dimer

Xiaoqing Liang[1,3], Xue Wu[1], Jun Hu[2,*], Jijun Zhao[1,*], Xiao Cheng Zeng[3,4*]

[1]*Key Laboratory of Materials Modification by Laser, Ion and Electron Beams (Dalian University of Technology), Ministry of Education, Dalian 116024, China*

[2]*College of Physics, Optoelectronics and Energy, Soochow University, Suzhou, Jiangsu 215006, China*

[3]*Department of Chemistry, University of Nebraska, Lincoln, NE 68588, USA*

[4]*Collaborative Innovation Center of Chemistry for Energy Materials, University of Science and Technology of China, Hefei 230026, China*

**Abstract**

Exploring giant magnetic anisotropy in small magnetic nanostructures is of both fundamental interest and technological merit for information storage. To prevent spin flipping at room temperature due to thermal fluctuation, large magnetic anisotropy energy (MAE) over 50 meV in magnetic nanostructure is desired for practical applications. We chose one of the smallest magnetic nanostructures — $Ir_2$ dimer, to investigate its magnetic properties and explore possible approach to engineer the magnetic anisotropy. Through systematic first-principles calculations, we found that the $Ir_2$ dimer already possesses giant MAE of 77 meV. We proposed an effective way to enhance the MAE of the $Ir_2$ dimer to 223 ~ 294 meV by simply attaching a halogen atom at one end of the Ir-Ir bond. The underlying mechanism for the record high MAE is attributed to the modification of the energy diagram of the $Ir_2$ dimer by the additional halogen-Ir bonding, which alters the spin-orbit coupling Hamiltonian and hence the magnetic anisotropy. Our strategy can be generalized to design other magnetic molecules or clusters with giant magnetic anisotropy.

[*]Corresponding authors. Email: jhu@suda.edu.cn (J. H); zhaojj@dlut.edu.cn (J.Z); xzeng1@unl.edu (X.C.Z).

The continuous miniaturization of the spintronics devices for modern technologies such as magnetic data storage would eventually reach the ultimate length scale (i.e. one to a few atoms) [1-6]. Recently, reading and writing quantum magnetic states in magnetic nanostructures with only a few transition-metal atoms were achieved by several experimental groups [1-7]. These investigations demonstrate the fascinating possibility to utilize magnetic nanostructure and even singe atom in the nanometer-scale spintronics devices. In this realm, the magnetic anisotropy of a magnetic nanostructure is the critical factor because it prevents the random spin reorientation induced by thermal fluctuation. Therefore, large magnetic anisotropic energy (MAE) is desired in magnetic nanostructures that serve as the building blocks of spintronics devices. Typical magnetic nanostructures based on 3$d$ transition-metal (TM) atoms have MAEs of only a few meV which corresponds to a blocking temperature ($T_B$) under 50 K, implying that their magnetic states are stable only at very low temperature [2,8]. For practical applications of the magnetic nanostructures at room temperature, large MAEs up to about 30 ~ 50 meV are necessary.

The magnetic anisotropy of a magnetic nanostructure originates from the spin-orbit coupling (SOC) effect. By analysis of the SOC Hamiltonian based on the second-order perturbation theory, Wang *et al.* expressed the MAE as the competition between the angular momentum matrix elements $\langle L_z \rangle$ and $\langle L_x \rangle$ [9,10]:

$$MAE \approx \xi^2 \sum_{u\alpha,o\beta} (2\delta_{\alpha\beta} - 1) \left[ \frac{\langle u\alpha | L_z | o\beta \rangle^2}{E_{u\alpha} - E_{o\beta}} - \frac{\langle u\alpha | L_x | o\beta \rangle}{E_{u\alpha} - E_{o\beta}} \right] \qquad (1)$$

Here ξ is the SOC constant; $E_{u\alpha}$ and $E_{o\beta}$ are the energy levels of the unoccupied states with spin $\alpha$ ($|u\alpha\rangle$) and the occupied states with spin $\beta$ ($|o\beta\rangle$), respectively. Therefore, the keys to achieve large MAE are: (i) large SOC constant ξ which exists in heavy elements such as 5$d$ TM elements; (ii) specific energy diagram to reduce the denominator in Eq. (1) that can be realized by appropriate ligand field. For example, a large MAE of 9 meV in a single Co atom was induced by placing it on Pt(111) substrate [11], while the MAE per atom in hcp Co solid is only 0.045 meV [12]. On the other hand, a giant MAE of 60 meV for Co adatom on MgO(001) surface was observed in a recent experiment, and such giant MAE was attributed to the special ligand field of the substrate [4]. Interestingly, combining the effects of the ligand field, orbital multiplet, and large SOC constants, even larger MAEs of 110 meV and 208 meV for Ru and Os adatom on MgO(001) surface were predicted based on density functional theory (DFT) calculations [13]. Recently, attempts to attain large MAEs with 4$f$ TM elements such as Dy, Ho and Er on substrate were also made [14-16]. Despite of the large SOC constants for these 4$f$ TM elements, the resulting MAEs are relatively small, i.e., 21.4 meV for Dy on graphene, 5.3 meV and 3.9 meV for Er and Ho atom on Pt(111), respectively.

In addition to the single TM adatom on specific substrate, TM dimers are of particular interest [17-21] due to their special symmetry. A homo-nuclear TM dimer is rotationally invariant around the molecular axis. Consequently, its magnetic anisotropy would arise at the first-order perturbation treatment of SOC interaction and therefore can be anomalously large compared to most other magnetic nanostructures [17]. In fact, appreciable MAEs of 30 ~ 70 meV were predicted theoretically for some free-standing TM dimers (positive MAE means easy axis parallel to the dimer axis) [20,22,23].

Intuitively, the energy diagram of a given TM dimer can be modified effectively by chemical functionalization, which can in turn affect the magnetic anisotropy as clearly shown in Eq. (1). Therefore, it is tempting to explore feasible ways to enhance the magnetic anisotropy of TM dimers. Besides placing the TM dimer on substrate, one possible tactic is to attach a light non-metal atom which can form strong chemical bond with the metal atom(s) (thus affect the energy diagram), yet still retain the spin moment of the entire cluster. Here, we choose Ir$_2$ dimer as a prototype to explore the possibility of engineering its magnetic anisotropy since Ir$_2$ possesses the largest MAE (70 meV) among free-standing homonuclear TM dimers [20]. Our first-principles calculations demonstrate that record high MAE up to 294 meV can be achieved in the Ir$_2$ dimer functionalized with a halogen atom (F, Cl, Br, I). To our knowledge, such a giant MAE is greater than any MAE value reported for small magnetic clusters in the literature, and is thus of great potential for application in spintronics devices.

We constructed various structures for $Ir_2X$ (X is non-metal anion atom, including F, Cl, Br, I) with X atom being located at different positions, including linear and noncollinear configurations to guarantee a complete sampling of the potential energy surface of the clusters. The structural relaxations and electronic structure calculations were carried out with spin-polarized local density approximation (LDA) in the Ceperley-Alder scheme [24,25], as implemented in OpenMX package [26]. To determine the MAEs and orbital magnetic moments, self-consistent calculations with inclusion of SOC effect were performed. The MAEs of the linear configurations for both $Ir_2$ dimer and $Ir_2X$ trimers are defined as the difference between the total energies with the magnetization parallel ($E(//)$) and perpendicular ($E(\perp)$) to the Ir-Ir bond direction: $MAE = E(//) - E(\perp)$.

Firstly, we calculate the structural and magnetic properties of $Ir_2$ dimer as given in Table 1. The equilbrium Ir-Ir bond length is 2.24 Å, slightly shorter than the experimental value (2.27 Å) [27], but close to a previously computed value based on the DFT/PW91 method (2.22 Å) [20]. The binding energy is 5.4 eV, significantly larger than the experimental mesurement (3.5 eV) [28] but agrees with previous calculation based on the LDA method with LanL2DZ-BSSE basis set correction (5.5 eV) [29]. It is known that LDA tends to lead over-binding for metal clusters. Qualitatively, the large binding energy reflects the strong interaction between the two Ir atoms. The computed spin moment ($M_S$) is 4 $\mu_B$ without considering the SOC effect. With considering the SOC, the $M_S$ reduces slightly to 3.86 $\mu_B$ while a large orbital moment ($M_L$) of 2.06 $\mu_B$ is induced, in line with previous calculation [20]. The MAE is 77 meV with easy axis parallel to the molecular axis. The present MAE is slightly larger than that reported previously (70 meV) [20], probably due to different choice of DFT methods (e.g., exchange-correlation functional and basis set).

Table 1. The ground state geometry, interatomic distance along easy-axis (*d*), total binding energy (E$_b$), spin moment (M$_S$), orbital moment (M$_L$), and magnetic anisotropy energy (MAE) for Ir$_2$ and Ir$_2$X (X=F, Cl, Br, I) clusters. The interatomic distances are shown as Ir-Ir (Ir-X) bond lengths for Ir$_2$X. The positive MAE indicates easy axis parallel to the molecular axis. The record high MAE value is highlighted in bold.

|  | Geometry | *d* (Å) | E$_b$ (eV) | M$_S$ (μ$_B$) | M$_L$ (μ$_B$) | MAE (meV) |
|---|---|---|---|---|---|---|
| Ir$_2$ | 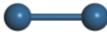 | 2.24 | 5.4 | 3.86 | 2.06 | 77 |
| Ir$_2$F | 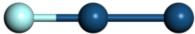 | 2.23 (1.87) | 10.7 | 3.02 | 0.98 | 232 |
| Ir$_2$Cl | 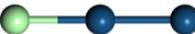 | 2.24 (2.19) | 9.5 | 3.04 | 0.98 | 223 |
| Ir$_2$Br | 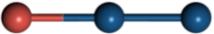 | 2.24 (2.31) | 9.3 | 3.03 | 0.98 | **294** |
| Ir$_2$I | 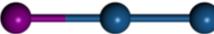 | 2.24 (2.49) | 8.8 | 3.01 | 0.99 | 228 |

To understand the origin of the novel magnetic characteristic of Ir$_2$, the energy diagram of its molecular orbitals is plotted in Figure 1(a). Due to the rotational symmetry of Ir$_2$ dimer, the Ir-5*d* orbitals split into three groups: $d_{xy/x^2-y^2}$, $d_{xz/yz}$ and $d_{z^2}$. The basal plane of $d_{xy/x^2-y^2}$ is perpendicular to the Ir-Ir bond, and those of $d_{xz/yz}$ and $d_{z^2}$ are crossing the Ir-Ir bond. Consequently, the interaction between the two Ir atoms results in three types of hybridizations: $d_{xy/x^2-y^2} - d_{xy/x^2-y^2}$, $d_{xz/yz} - d_{xz/yz}$, and $d_{z^2} - d_{z^2}$. The corresponding bonding and antibonding states (i.e. six molecular orbitals in each spin channel) can be thus notated as $\delta_d/\delta_d^*$, $\pi_d/\pi_d^*$ and $\sigma_d/\sigma_d^*$. Based on the spatial wave functions in Figure 1(b) and the spin-polarized projected density of states (PDOS) in Figure 2(a) of these molecular orbitals, we identified the characters of the energy levels and marked them in the energy diagram in Figure 1(a). Clearly, the hybridization between the two $d_{z^2}$ orbitals of the two Ir atoms is much stronger than those between the other orbitals. This leads to large separation in energy between $\sigma_d$ and $\sigma_d^*$. The interaction between the two $d_{xy/x^2-y^2}$ orbitals is weakest, thus the separation of their energy levels is smallest. In majority spin channel, all these molecular orbitals are occupied. In minority spin channel, the doubly degenerate $\delta_d^*$ orbital is half occupied, and $\pi_d^*$ and $\sigma_d^*$ are unoccupied. Therefore, the electronic configuration of these molecular orbitals is $(\sigma_d)^2(\pi_d)^4(\delta_d)^4(\delta_d^*)^3(\pi_d^*)^2(\sigma_d^*)^1$. On the other hand, the interaction between the Ir-$s$ orbitals is also strong, and the antibonding-state molecular orbitals ($\sigma_s^*$) in both spin channels are unoccupied, resulting in an electronic configuration $(\sigma_s)^2(\sigma_s^*)^0$. In addition, there is moderate hybridization between $d_{z^2}$ and $s$ orbitals, as shown in Figure 2(a). Note that the electronic configuration of an isolated Ir atom is $(5d)^7(6s)^2$. Accordingly, we can conclude that each Ir-6$s$ orbital donates one electron to Ir-5$d$ orbitals. As a result, the total spin moment of $Ir_2$ is 4 $\mu_B$, contributed by $\delta_d^*$ (1 $\mu_B$), $\pi_d^*$ (2 $\mu_B$), and $\sigma_d^*$ (1 $\mu_B$), respectively. With inclusion of the SOC effect, the doubly degenerate $\delta_d^*$ orbital in the minority spin channel splits significantly (by ~0.9 eV) for the easy-axis magnetization, while the hard-axis magnetization does not change the energy diagram much. Furthermore, the spin and orbital moments possess large anisotropy. As seen in Table S1 of Supplemental Material, the spin moments along the easy and hard axis are 3.86 and 3.28 $\mu_B$, while the corresponding orbital moments are 2.06 and 1.16 $\mu_B$, respectively. Intuitively, the combined effects of large SOC splitting of $\delta_d^*$ in the minority spin channel and the large anisotropy of the spin and orbital moments are responsible for the large MAE of $Ir_2$ [17].

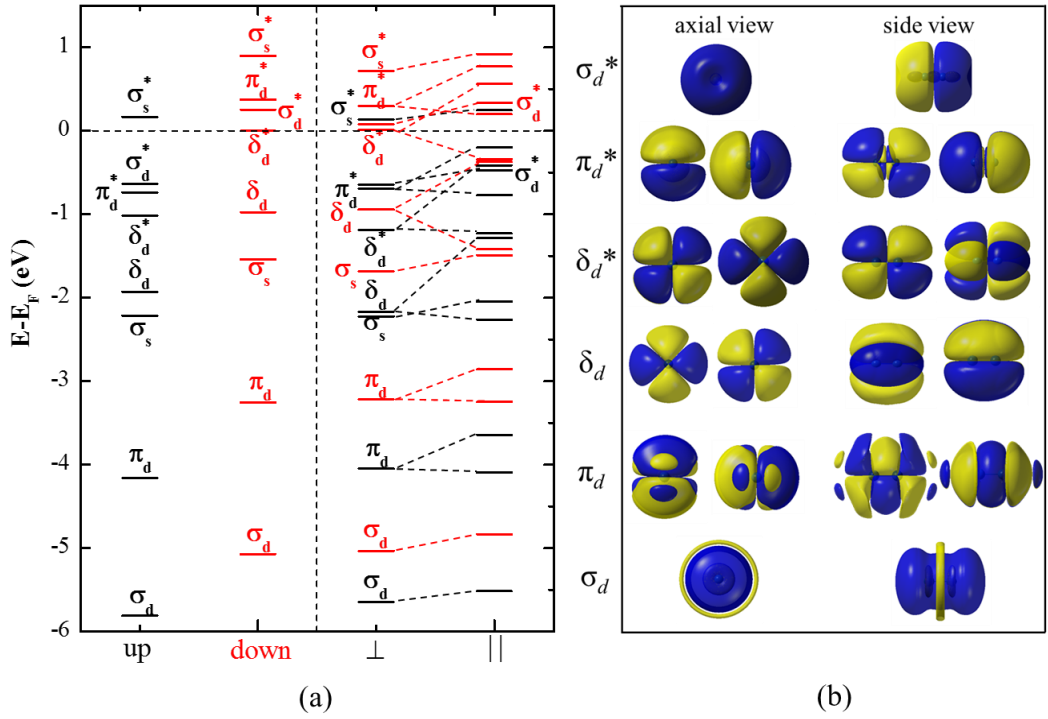

Figure 1. (a) Energy diagram of the molecular orbitals of the Ir$_2$ dimer without (left panel) and with (right panel) considering the SOC effect. Black refers to majority spin-up channel while red refers to minority spin-down channel. Both the magnetizations perpendicular (⊥) and parallel (∥) (easy axis) to the Ir-Ir bond were considered. SOC induced splitting of degenerated orbitals are denoted by the branched dashed lines. (b) The spatial wave functions of the molecular orbitals without considering the SOC effect. The paratactic orbitals are energetically degenerate. Yellow and blue correspond to different phases of the molecular wave functions, and the cutoff value of the isosurfaces is ±0.02 a.u.

To further elucidate the underlying mechanism of the large MAE of Ir$_2$, we estimated the MAE of Ir$_2$ by using Eq. (1) to distinguish the contributions from different angular momentum matrices to the MAE. For convenience, we divide the total MAE into three parts: $MAE = MAE_{uu} + MAE_{dd} + MAE_{ud}$. These three terms are due to the coupling between the majority spin states ($uu$), the minority spin states ($dd$), and cross-spin states ($ud$), respectively. Since the $s$ orbital is not involved in the SOC, the MAE is determined by the molecular orbitals from the Ir-5$d$ states only. From the energy diagram in Figure 1(a) and the PDOS in Figure 2(a), the Ir-5$d$ related molecular orbitals in the majority spin channel are completely occupied, thus $MAE_{uu}$ is negligible. For $MAE_{dd}$ and $MAE_{ud}$, the nonzero contributions from the angular moment matrix elements in Eq. (1) are plotted in Figure 2(b). Clearly, for $MAE_{dd}$, the contribution from $L_z$ is much larger than that from $L_x$, resulting in a positive $MAE_{dd}$ (~ $10\xi^2$). For cross-spin coupling, $L_z$ still contributes more to $MAE_{ud}$ than $L_x$, but this leads to a negative $MAE_{ud}$ (~ $-3\xi^2$) due to the negative sign in Eq. (1). Nevertheless, the absolute magnitude of $MAE_{dd}$ is larger than that of $MAE_{ud}$, resulting in a positive total MAE (about $7\xi^2$) for Ir$_2$. Furthermore, it can be seen from Figure 2(b) that both the sign and magnitude of $MAE_{dd}$ and $MAE_{ud}$ are dominated by the $L_z$ matrix elements that are related to the $\delta_d$ and $\delta_d^*$ orbitals: $\langle\delta_d,\downarrow|L_z|\delta_d^*,\downarrow\rangle$ (~$17\xi^2$), $\langle\delta_d,\uparrow|L_z|\delta_d^*,\downarrow\rangle$ (~$8\xi^2$) and $\langle\delta_d^*,\uparrow|L_z|\delta_d^*,\downarrow\rangle$ (~$16\xi^2$). Therefore, the MAE of the Ir$_2$ dimer is mainly attributed to the SOC effect associated with the $d_{xy/x^2-y^2}$ orbitals.

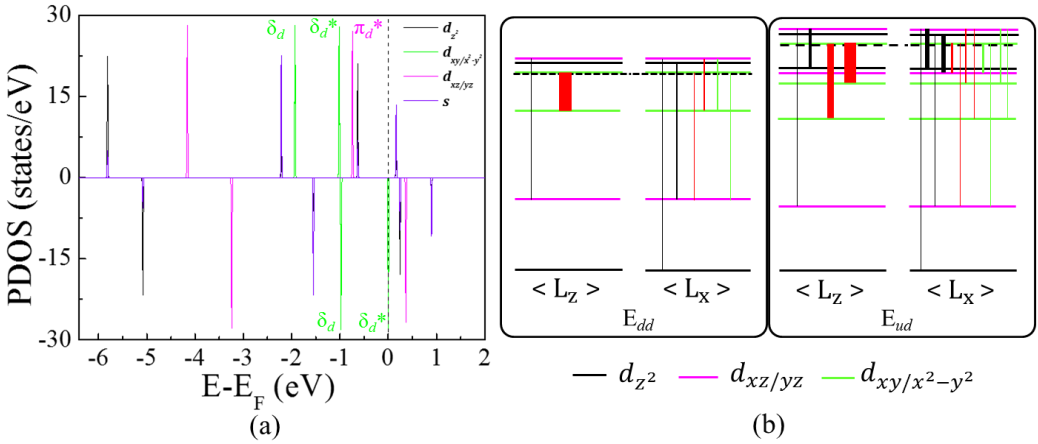

Figure 2. (a) Projected density of states (PDOS) of the $s$ and $d$ orbitals of Ir$_2$ dimer. The vertical dashed line marks the Fermi level (E$_F$). (b) Sketches of the nonzero contribution from each pair of molecular orbitals to the MAE between minority spin states (left panel marked as E$_{dd}$) and cross-spin states (right panel marked as E$_{ud}$). The horizontal dashed line shows the position of E$_F$ for each case. The thickness of each vertical line scales with the magnitude of the corresponding SOC matrix element and colors are used to distinguish different contribution from a pair of states.

According to the discussions above and Eq. (1), if an effective way to modify the energy diagram can be found, then the contribution from each pair of the molecular orbitals may be revised, thereby altering the total MAE. One such an effective tactic is by attaching an additional anion atom to $Ir_2$ to tailor the interaction between the two Ir atoms and the magnetic property of $Ir_2$. To this end, we examined a series of non-metal anion atom X, including C, Si, N, P, O, S, and halogen elements, to construct $Ir_2X$ clusters. As shown in Tables S2 and S3, there are three types of possible equilibrium structures: (I) linear chain with X atom at one end; (II) isosceles triangle with X atom over the middle point of the Ir-Ir bond; (III) linear chain with X atom at the middle point. We obtained the ground-state structure and MAE for all the $Ir_2X$ clusters and found that only the halogen atoms can result in huge MAEs. Hereafter, we only take the halogen elements as a prototype to discuss the strategy of chemical engineering the MAE of transition metal dimer.

For all halogen elements X (X = F, Cl, Br, I), the $Ir_2X$ clusters prefer type-I structure (see Table S3 in Supplemental Material). As seen from Table 1, the Ir-Ir bond length in $Ir_2X$ is slightly changed by X, while the Ir-X bond length increases monotonically with X for X = F to I due to the increasing atomic radius. Moreover, the binding energies of the $Ir_2X$ trimers are about twice of that of $Ir_2$, indicating high structural stability of the entire trimer and strong binding between Ir and X atoms. Consequently, the interaction between the two Ir atoms is significantly modified, as manifested by the PDOS of $Ir_2F$, as an example, in Figure 3(a). It can be seen that the F-$2p$ orbitals hybridize strongly with the $d_{xz/yz}$ and $d_{z^2}$ orbitals of the Ir1 atom (bonded with X). Compared to those of $Ir_2$, the bonding states of both $d_{xz/yz}$ and $d_{z^2}$ orbitals of the Ir1 atom shift downward by about 0.3 eV in the majority spin channel and 0.6 eV in the minority spin channel, while the antibonding states of $d_{xz/yz}$ and $d_{z^2}$ shift upward by about 0.7 eV and 0.3 eV, respectively. As a result, the hybridization between the two Ir atoms through the $d_{xz/yz}$ and $d_{z^2}$ orbitals is markedly weakened. The energy levels of the $d_{xz/yz}$ and $d_{z^2}$ orbitals of the Ir2 atom (at the other end of Ir$_2$X) shift upwards, and the energy separations between the corresponding bonding and antibonding states become narrower. Meanwhile, some electrons are transferred from the $d_{xz/yz}$ and $d_{z^2}$ orbitals of the Ir atoms to the F atom, as shown by the inset in Figure 3(a). On the other hand, the interaction between the $d_{xy/x^2-y^2}$ orbitals of the Ir atoms is not affected much, but slightly weakened only. Therefore, the $\delta_d$ and $\delta_d^*$ are similar with those of Ir$_2$ but their energy separation is narrowed by about 0.24 eV [see Figure 3(a)]. At the same time, the exchange splitting increases by about 0.52 eV.

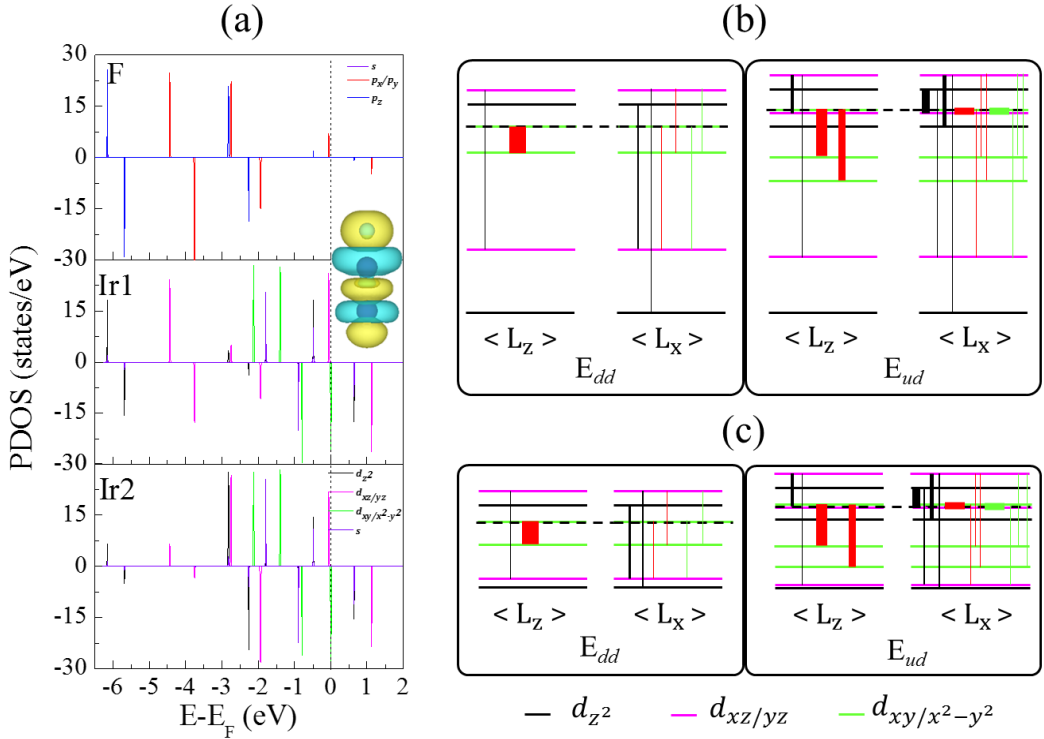

Figure 3. (a) Projected density of states (PDOS) of Ir$_2$F, $s$ and $p$ orbitals of F atom and $s$ and $d$ orbitals of two Ir atoms (Ir1 is the neighboring atom of F atom and Ir2 is at the end of Ir$_2$F) are separately shown in the panels. The insets are the corresponding charge redistribution. The yellow and blue areas refer to the electron accumulation and depletion, respectively, with isosurface value of 6×10$^{-3}$ e/Å$^3$. The vertical dashed lines mark the Fermi level (E$_F$). (b, c) Sketches of the nonzero contribution from each pair of molecular orbitals to the MAE between minority spin states (left panel marked as $E_{dd}$) and cross-spin states (right panel marked as $E_{ud}$) for the two Ir atoms of Ir$_2$F. The horizontal dashed line shows the position of Fermi level $E_F$ for each case. The width of each vertical line scales with the magnitude of the corresponding SOC matrix element and the colors are used to distinguish different contribution from a pair of states.

The magnetic moments are also significantly modified by attaching X atom to the Ir$_2$ dimer. Without considering SOC, the total spin moments of these Ir$_2$X trimers are all 5 μ$_B$, larger than that of Ir$_2$ by 1 μ$_B$. According to the Milliken population analysis, the Ir$_2$ dimer donates the electron of $\delta_d^*(\downarrow)$ to the halogen atom X. Thus, the spin moment is contributed by $\delta_d^*$ (2 μ$_B$), $\pi_d^*$ (2 μ$_B$), and $\sigma_d^*$ (1 μ$_B$). However, the SOC effect notably reduces M$_S$ of Ir$_2$X to 3 μ$_B$ (see Table 1), which is even smaller than that of Ir$_2$ by 1 μ$_B$. On the contrary, the M$_S$ of Ir$_2$ itself is only slightly altered by the SOC. From the PDOS without considering SOC (Fig. 3(a)), we can see that both $\pi_d^*(\uparrow)$ and $\delta_d^*(\downarrow)$ are very close to the $E_F$, which is pinned within the small gap (61 meV) between them. With considering SOC (axial magnetization), both $\pi_d^*(\uparrow)$ and $\delta_d^*(\downarrow)$ split into two orbitals, with splitting of the corresponding energy levels being as large as 498 meV and 896 meV, respectively (see Fig. S1 in Supplemental Material). Consequently, the higher energy level stemmed from $\pi_d^*(\uparrow)$ shifts upward above the $E_F$, while the lower energy level stemmed from $\delta_d^*(\downarrow)$ shifts downward below the $E_F$. One electron of $\pi_d^*(\uparrow)$ in majority spin channel transfers to $\delta_d^*(\downarrow)$ in minority spin channel, resulting in reduction of spin moment by 1 μ$_B$, i.e. the distribution of the spin moment changes to $\delta_d^*$ (1 μ$_B$), $\pi_d^*$ (1 μ$_B$), and $\sigma_d^*$ (1 μ$_B$). In addition, the orbital moments of Ir$_2$X trimers are about 1 μ$_B$, which is about half of that of Ir$_2$. As shown in Table S1 of Supplemental Material, both the local spin moment and orbital moment on Ir1 atom are smaller than those on Ir2 atom due to the hybridization between Ir1 and X atoms. It is worthy of noting that the orbital moments in Ir$_2$X trimers show very little anisotropy between the easy and hard axis, despite of the enhanced anisotropy of spin moments with respect to Ir$_2$.

Remarkably, the MAEs of all Ir$_2$X trimers are largely enhanced (see Table 1). Among them, Ir$_2$Cl entails the least MAE yet the MAE value is still as large as 223 meV; Ir$_2$Br has the largest MAE of 294 meV; and the MAEs of the other two trimers are about 230 meV. All these MAE values are larger than the highest MAE values reported in the literature for small magnetic clusters. The largest MAE value we are aware of is 208 meV for the Os adatom on MgO(001) [13].

To confirm our view of the underlying mechanism for the extraordinary enhancement of MAE due to functionalization by halogen atoms, we extracted the energy levels from the PDOS in Figure 3(a) and estimated the MAEs of Ir$_2$F using Eq. (1), as plotted in Fig. 3(b) and 3(c). From the nonzero contributions of the angular momentum shown in Fig. 3(b) and 3(c), the main contributions stem from the matrix elements of $L_z$ ($\langle \delta_d, \downarrow | L_z | \delta_d^*, \downarrow \rangle$, $\langle \delta_d, \uparrow | L_z | \delta_d^*, \downarrow \rangle$, $\langle \delta_d^*, \uparrow | L_z | \delta_d^*, \downarrow \rangle$) and $L_X$ ($\langle \pi_d^*, \uparrow | L_x | \delta_d^*, \downarrow \rangle$) of both Ir atoms. Compared to Ir$_2$ (see Table S4 in Supplemental Material), the positive contribution of $\langle \delta_d, \downarrow | L_z | \delta_d^*, \downarrow \rangle$ increases weakly and the negative contributions of $\langle \delta_d, \uparrow | L_z | \delta_d^*, \downarrow \rangle$ and $\langle \delta_d^*, \uparrow | L_z | \delta_d^*, \downarrow \rangle$ are slightly reduced. Interestingly, the contribution from $\langle \pi_d^*, \uparrow | L_x | \delta_d^*, \downarrow \rangle$ is minor in Ir$_2$ due to the relatively large energy separation between the corresponding orbitals. For Ir$_2$F, however, the energy levels of $\pi_d^*(\uparrow)$ and $\delta_d^*(\downarrow)$ become very close [~61 meV; see Fig. 3(a)], which results in huge contribution to MAE from $\langle \pi_d^*, \uparrow | L_x | \delta_d^*, \downarrow \rangle$. The final estimated *MAE$_{dd}$* and *MAE$_{ud}$* are 13$\xi^2$ and 44$\xi^2$, respectively, both being larger than the corresponding values for Ir$_2$ (10$\xi^2$ and −3$\xi^2$). Consequently, the total MAE of Ir$_2$F increases dramatically to 232 meV, about three times of that of Ir$_2$ (77 meV). In fact, for all Ir$_2$X trimers considered here, the term $\langle \pi_d^*, \uparrow | L_x | \delta_d^*, \downarrow \rangle$ dominates the MAE (see Table S4 in Supplemental Material) because they all have similar energy diagrams (see Fig. S2 in Supplemental Material).

In conclusion, our first-principles calculations demonstrate that the MAE of the Ir$_2$ dimer can be significantly enhanced up to a record high value of 294 meV by attaching a Br atom at one end of the Ir-Ir bond. Analysis of the energy levels and the matrix elements of the SOC Hamiltonian show that the $d_{xy/x^2-y^2}$ and $d_{xz/yz}$ orbitals are mainly responsible for the record high MAE value. More specifically, the halogen atoms with strong electronegativity lead to the stable linear configuration for the Ir$_2$X trimer, and induce large spin moments. The strategy of functionalization of TM-based cluster introduces a new synthetic approach to chemically engineering the magnetic anisotropy of TM clusters towards future-generation magnetic information storages using single or a few atoms per bit.

**Acknowledgements**

This work was supported by the National Natural Science Foundation of China (11574040, 11574223), the Natural Science Foundation of Jiangsu Province (BK20150303), the Fundamental Research Funds for the Central Universities of China (DUT16-LAB01, DUT17LAB19), and the Supercomputing Center of Dalian University of Technology. J.H. thanks the Jiangsu Specially-Appointed Professor Program of Jiangsu Province. X.C.Z. was supported by a State Key R&D Fund of China (2016YFA0200604) to USTC and University of Nebraska Holland Computing Center.

## References

[1] A. A. Khajetoorians, J. Wiebe, B. Chilian, and R. Wiesendanger, Science **332**, 1062 (2011).

[2] S. Loth, S. Baumann, C. P. Lutz, D. M. Eigler, and A. J. Heinrich, Science **335**, 196 (2012).

[3] A. A. Khajetoorians *et al.*, Science **339**, 55 (2013).

[4] I. G. Rau *et al.*, Science **344**, 988 (2014).

[5] T. Miyamachi *et al.*, Nature **503**, 242 (2013).

[6] M. Steinbrecher, A. Sonntag, M. dos Santos Dias, M. Bouhassoune, S. Lounis, J. Wiebe, R. Wiesendanger, and A. Khajetoorians, Nat. Commun. **7**, 10454 (2016).

[7] F. D. Natterer, K. Yang, W. Paul, P. Willke, T. Choi, T. Greber, A. J. Heinrich, and C. P. Lutz, Nature **543**, 226 (2017).

[8] F. Meier, L. Zhou, J. Wiebe, and R. Wiesendanger, Science **320**, 82 (2008).

[9] D.-s. Wang, R. Wu, and A. J. Freeman, Phy. Rev. B **47**, 14932 (1993).

[10] J. Hu and R. Wu, Phys. Rev. Lett. **110**, 097202 (2013).

[11] P. Gambardella *et al.*, Science **300**, 1130 (2003).

[12] D. Weller and A. Moser, IEEE Transactions on Magnetics **35**, 4423 (1999).

[13] X. Ou, H. Wang, F. Fan, Z. Li, and H. Wu, Phys. Rev. Lett. **115**, 257201 (2015).

[14] R. Baltic, M. Pivetta, F. Donati, C. Wäckerlin, A. Singha, J. Dreiser, S. Rusponi, and H. Brune, Nano Lett. **16**, 7610 (2016).

[15] F. Donati, A. Singha, S. Stepanow, C. Wäckerlin, J. Dreiser, P. Gambardella, S. Rusponi, and H. Brune, Phys. Rev. Lett. **113**, 237201 (2014).

[16] F. Donati *et al.*, Science **352**, 318 (2016).

[17] T. O. Strandberg, C. M. Canali, and A. H. MacDonald, Nat. Mater. **6**, 648 (2007).

[18] T. O. Strandberg, C. M. Canali, and A. H. MacDonald, Phy. Rev. B **77**, 174416 (2008).

[19] D. Fritsch, K. Koepernik, M. Richter, and H. Eschrig, J. Comput. Chem. **29**, 2210 (2008).

[20] P. Błoński and J. Hafner, Phy. Rev. B **79**, 224418 (2009).

[21] B. Piotr and H. Jürgen, J. Phys.: Condens. Matter **26**, 146002 (2014).

[22] H. K. Yuan, H. Chen, A. L. Kuang, B. Wu, and J. Z. Wang, J. Phys. Chem. A **116**, 11673 (2012).

[23] P. Wang, X. Jiang, J. Hu, X. Huang, and J. Zhao, J. Mater. Chem. C **4**, 2147 (2016).

[24] D. M. Ceperley and B. Alder, Phys. Rev. Lett. **45**, 566 (1980).

[25] J. P. Perdew and A. Zunger, Physical Review B **23**, 5048 (1981).

[26] H. K. T. Ozaki, J. Yu, M.J. Han, N. Kobayashi, M. Ohfuti, F. Ishii, et al. , User's manual of OpenMX version 3.7.

[27] M. D. Morse, Chem. Rev. **86**, 1049 (1986).

[28] A. Miedema and K. A. Gingerich, J. Phys. B: At. Mol. Phys. **12**, 2081 (1979).

[29] A. Posada-Borbón and A. Posada-Amarillas, Chem. Phys. Lett. **618**, 66 (2015).

# Supplemental Material

Table S1. On-site spin moments and orbital moments (in μB) of Ir and X atoms along easy magnetization direction ($M_S^e$ and $M_L^e$) and hard magnetization direction ($M_S^h$ and $M_L^h$), respectively. Ir1 is the neighboring atom of X atom and Ir2 atom is located at the end of Ir$_2$X (X = F, Cl, Br, I).

|  | $M_S^e$ | | | $M_L^e$ | | | $M_S^h$ | | | $M_L^h$ | | |
|---|---|---|---|---|---|---|---|---|---|---|---|---|
|  | Ir1/Ir2 | X | Total | Ir1/Ir2 | X | Total | Ir1/Ir2 | X | Total | Ir1/Ir2 | X | Total |
| Ir$_2$ | 1.93/1.93 | -- | 3.86 | 1.03/1.03 | -- | 2.06 | 1.64/1.64 | -- | 3.28 | 0.58/0.58 | -- | 1.16 |
| Ir$_2$F | 1.40/1.55 | 0.08 | 3.02 | 0.40/0.68 | –0.09 | 0.98 | 0.85/0.91 | 0.03 | 1.79 | 0.43/0.46 | –0.02 | 0.86 |
| Ir$_2$Cl | 1.36/1.57 | 0.11 | 3.04 | 0.48/0.66 | –0.15 | 0.98 | 0.81/0.96 | 0.03 | 1.79 | 0.43/0.47 | –0.03 | 0.87 |
| Ir$_2$Br | 1.37/1.58 | 0.09 | 3.03 | 0.49/0.66 | –0.16 | 0.98 | 1.62/1.80 | 0.12 | 3.55 | 0.49/0.52 | –0.01 | 0.99 |
| Ir$_2$I | 1.38/1.58 | 0.05 | 3.01 | 0.52/0.66 | –0.19 | 0.99 | 0.79/0.98 | 0.01 | 1.78 | 0.45/0.43 | –0.01 | 0.87 |

Table S2. The atomic configurations and energy differences (ΔE) of Ir$_2$ dimer bonded with a carbon, silicon, pnictogen, or chalcogen atom. The spin moment (M$_S$) without inclusion of the SOC, easy magnetization direction (easy-axis) and the magnetic anisotropy energy (MAE) are listed for the ground-state structure. All results are calculated by using LDA method implemented in OpenMX package.

| X↑ →Z | Configuration | ΔE (eV) | M$_S$ (μ$_B$) | easy-Axis | MAE(meV) |
|---|---|---|---|---|---|
| Ir$_2$C | 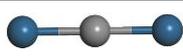 | 0 | 2 | Z | 10.9 |
|  | 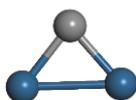 | 1.84 | 2 | -- | -- |
|  | 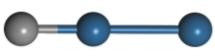 | 3.20 | 2 | -- | -- |
| Ir₂Si | 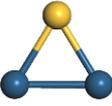 | 0 | 2 | *Y* | 8.8 |
| | 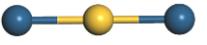 | 1.36 | 2 | -- | -- |
| | 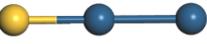 | 2.63 | 4 | -- | -- |
| Ir₂N | 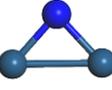 | 0 | 1 | *Z* | 2.2 |
| | 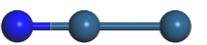 | 0.63 | 1 | -- | -- |
| | 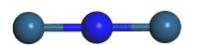 | 1.81 | 1 | -- | -- |
| Ir₂P | 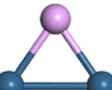 | 0 | 1 | *X* | 0 |
| | 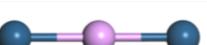 | 1.28 | 1 | -- | -- |
| | 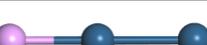 | 2.09 | 3 | -- | -- |
| Ir₂O | 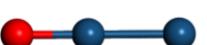 | 0 | 0 | -- | -- |
| | 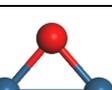 | 1.46 | 2 | -- | -- |
| | 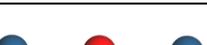 | 3.42 | 0 | -- | -- |
| Ir₂S | 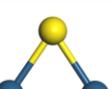 | 0 | 0 | -- | -- |
| | 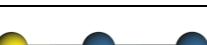 | 0.23 | 2 | -- | -- |
| | 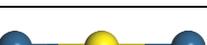 | 1.58 | 0 | -- | -- |

Table S3. The atomic configurations and energy differences (ΔE) of the metastable isomers (denoted as II and III) compared to the ground state structure (denoted as I), the bond lengths of Ir-Ir and Ir-X in the parentheses, and the total spin moments (M$_S$) without inclusion of the SOC.

| | Isomer | Configuration | $d$ (Å) | ΔE (eV) | M$_S$ ($\mu_B$) |
|---|---|---|---|---|---|
| Ir$_2$F | I | 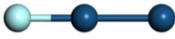 | 2.27 (1.91) | 0 | 5 |
| | II | 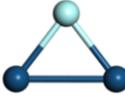 | 2.29 (2.19) | 1.69 | 5 |
| | III | 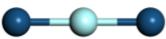 | -- (1.97) | 4.68 | 5 |
| Ir$_2$Cl | I | 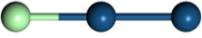 | 2.28 (2.23) | 0 | 5 |
| | II | 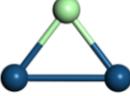 | 2.40 (2.26) | 1.01 | 3 |
| | III | 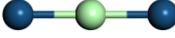 | -- (2.15) | 3.44 | 3 |
| Ir$_2$Br | I | 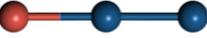 | 2.28 (2.35) | 0 | 5 |
| | II | 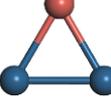 | 2.39 (2.36) | 0.75 | 3 |
| | III | 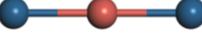 | -- (2.29) | 3.48 | 5 |
| Ir$_2$I | I | 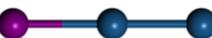 | 2.28 (2.53) | 0 | 5 |
| | II | 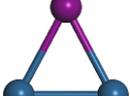 | 2.38 (2.53) | 0.36 | 3 |
| | III | 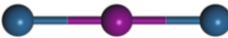 | -- (2.46) | 3.31 | 5 |

Table S4. The estimated contributions of MAE based on Eq.(1), including the total MAE$_{dd}$ and MAE$_{ud}$. The main contribution of MAE$_{dd}$ from the matrix element $L_z$ ($\langle \delta_d, \downarrow |L_z| \delta_d^*, \downarrow \rangle$) is listed separately, and those in the columns of MAE$_{ud}$ are the main contributions from the matrix elements of $L_Z$ ($\langle \delta_d, \uparrow |L_z| \delta_d^*, \downarrow \rangle$ and $\langle \delta_d^*, \uparrow |L_z| \delta_d^*, \downarrow \rangle$) and $L_X$ ($\langle \pi_d^*, \uparrow |L_x| \delta_d^*, \downarrow \rangle$). All values are in unit of constant $\xi^2$, where $\xi$ is the SOC constant.

|  | MAE$_{dd}$ | | MAE$_{ud}$ | | | |
|---|---|---|---|---|---|---|
|  | total | $\langle \delta_d, \downarrow \|L_z\| \delta_d^*, \downarrow \rangle$ | total | $\langle \delta_d, \uparrow \|L_z\| \delta_d^*, \downarrow \rangle$ | $\langle \delta_d^*, \uparrow \|L_z\| \delta_d^*, \downarrow \rangle$ | $\langle \pi_d^*, \uparrow \|L_x\| \delta_d^*, \downarrow \rangle$ |
| Ir$_2$ | 10 | 17 | −3 | −8 | −16 | 5 |
| Ir$_2$F | 13 | 20 | 44 | −13 | −9 | 49 |
| Ir$_2$Cl | 13 | 20 | 51 | −13 | −9 | 51 |
| Ir$_2$Br | 14 | 21 | 98 | −12 | −8 | 100 |
| Ir$_2$I | 12 | 21 | 220 | −12 | −8 | 222 |

Table S5. The spin moment (M$_S$) in $\mu_B$ without considering the SOC, and MAE values in meV computed based on the PBE method implemented in OpenMX, DMol$^3$, and VASP packages, and compared to LDA results by OpenMX.

|  | OpenMX | | | | VASP | | DMol$^3$ |
|---|---|---|---|---|---|---|---|
|  | M$_S$ (LDA) | MAE (LDA) | M$_S$ (GGA) | MAE (GGA) | M$_S$ (GGA) | MAE (GGA) | M$_S$ (GGA) |
| Ir$_2$ | 4 | 77 | 4 | 18 | 4 | 41 | 4 |
| Ir$_2$F | 5 | 232 | 5 | 211 | 5 | 202 | 5 |
| Ir$_2$Cl | 5 | 223 | 5 | 156 | 5 | 258 | 5 |
| Ir$_2$Br | 5 | 294 | 5 | 218 | 5 | 251 | 5 |
| Ir$_2$I | 5 | 228 | 5 | 141 | 5 | 230 | 5 |

Table S6. The Cartesian coordinates (Å) of Ir$_2$ and Ir$_2$X (X = F, Cl, Br, I) clusters in the ground state structures, based on LDA-CA method implemented in OpenMX.

| Cluster | | x | y | z |
|---|---|---|---|---|
| Ir$_2$ | Ir | 0.00000 | 0.00000 | −0.02035 |
| | Ir | 0.00000 | 0.00000 | 2.22035 |
| Ir$_2$F | F | 0.00000 | 0.00000 | 0.01478 |
| | Ir | 0.00000 | 0.00000 | 1.92503 |
| | Ir | 0.00000 | 0.00000 | 4.19082 |
| Ir$_2$Cl | Cl | 0.00000 | 0.00000 | 0.01015 |
| | Ir | 0.00000 | 0.00000 | 2.24440 |
| | Ir | 0.00000 | 0.00000 | 4.52141 |
| Ir$_2$Br | Br | 0.00000 | 0.00000 | 0.02207 |
| | Ir | 0.00000 | 0.00000 | 2.37409 |
| | Ir | 0.00000 | 0.00000 | 4.65362 |
| Ir$_2$I | I | 0.00000 | 0.00000 | 0.02347 |
| | Ir | 0.00000 | 0.00000 | 2.55427 |
| | Ir | 0.00000 | 0.00000 | 4.83460 |

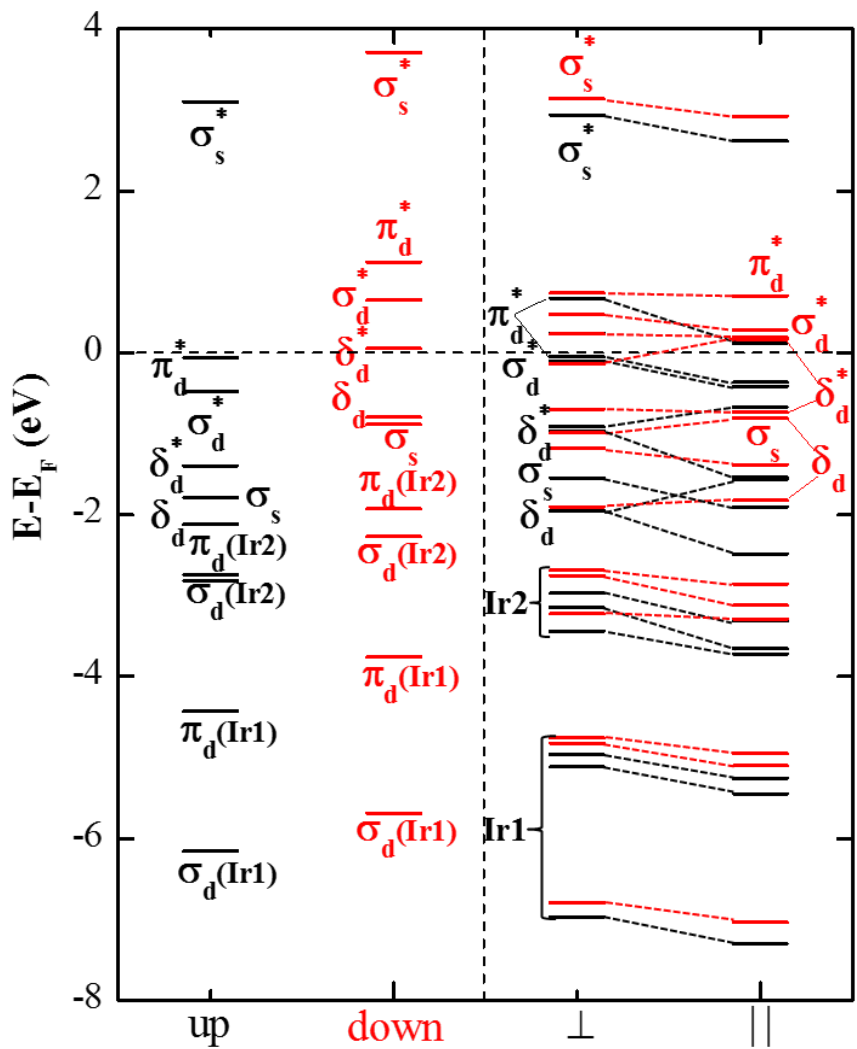

Figure S1．Energy diagram of the molecular orbitals of the Ir$_2$F without (left panel) and with (right panel) considering the SOC effect. Black refers to majority spin-up channel while red refers to minority spin-down channel. Both the magnetizations perpendicular (⊥) and parallel (∥) to the bond direction were considered. SOC induced splitting of degenerated orbitals are denoted by the branched dashed lines. The notations in the parentheses indicate that the corresponding energy levels are contributed dominantly by which Ir atom. The energy levels without such kind of notation are contributed by both Ir atoms comparably.

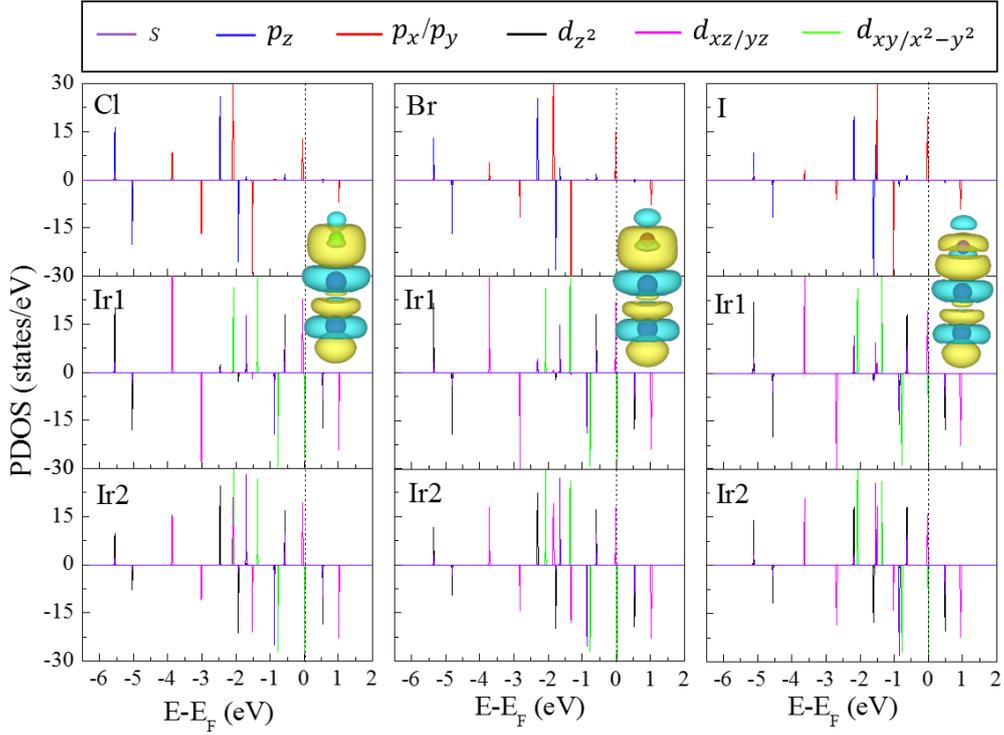

Figure S2. Projected density of states (PDOS) of $Ir_2X$ (X = Cl, Br, I), $s$ orbitals of all atoms, $p$ orbitals of X atom and $d$ orbitals of two Ir atoms (Ir1 is the neighboring atom of F atom and Ir2 is at the end of $Ir_2F$) are shown separately in the panels. The insets present the corresponding charge redistribution. The yellow and blue areas refer to the electron accumulation and depletion, respectively, with isosurface value of $6\times10^{-3}$ e/Å$^3$.

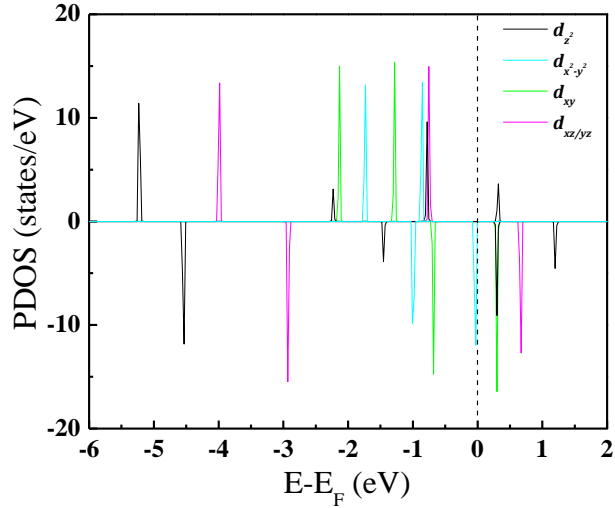

Figure S3. Projected density of states (PDOS) of the $d$ orbitals of $Ir_2$ dimer from computation executed in VASP. The vertical dashed line marks the Fermi level ($E_F$).

**Methods**

We used OpenMX software package which is based on density functional theories (DFT) [1,2], norm-conserving pseudopotentials [3-7], and pseudo-atomic localized basis functions[8,9]. The cutoff radii of the radial wave function were 9.0, 7.0, 9.0, 9.0 and 11.0 a.u. and the valence orbitals were $s^2p^2d^2f^1$, $s^3p^3d^2$, $s^3p^3d^2$, $s^4p^4d^3f^1$, and $s^3p^3d^3f^2$ for Ir, F, Cl, Br and I, respectively. The fully relativistic pseudopotentials [7] were used, and the cutoff energy was set to 300 Ry. The criteria for energy and force convergence were $10^{-7}$ Hartree and $10^{-4}$ Hartree/bohr, respectively.

**Justifications of different exchange-correlation functionals and different first-principles packages**

In order to justify whether the LDA is reliable to describe the magnetic properties including spin moment and magnetic anisotropy energy (MAE), we used PBE functional [10] within generalized gradient approximation (GGA), implemented in OpenMX, DMol$^3$ and VASP packages to reoptimize the structures of the Ir$_2$ and Ir$_2$X (X = F, Cl, Br, I) clusters, and compute their spin moments and MAEs accordingly. We carried out spin-polarized density functional theory calculations within GGA-PBE functional, implemented in VASP package [11,12]. The interaction between the valence electrons and ionic cores was described by the projector augmented wave (PAW) [13] method. Every individual clusters were placed in a simple cubic supercell of $20 \times 20 \times 20$ Å$^3$ to ensure sufficient separation between the periodic images. The energy cutoff for the plane-wave basis was set to 500 eV. During the self-consistent electronic structure calculations and geometry optimizations, the criteria for energy and force convergence were $10^{-6}$ eV and 10 meV/Å, respectively. We defined the easy magnetization orientation aligned parallel to the bond direction as positive MAE, while perpendicular to bond direction as negative MAE, which is convenient to compare MAEs by different methods. And also we used DMol$^3$ [14,15] to obtain the most stable magnetic state in different multiplicity, The GGA-PBE functional with basis-set of double numerical basis including *p*-polarization function are executed in DMol$^3$ package, the cutoff radius is 6 Å and the criteria for energy convergence is $10^{-7}$ Hartree during the selfconsistent electronic calculation.

For comparison, all the results calculated with different methods and first-principles packages are summarized in Table S5. First, the total spin moments for each cluster are consistent. However, the MAEs of Ir$_2$X calculated with GGA functional using either OpenMX or VASP are generally smaller than the values calculated with LDA in OpenMX, but still large enough to prevail most reported values in the literature. To explore the origin of MAE of a TM cluster, accurate energy levels are needed. From the calculations using different packages and different functionals, we found that only the LDA in OpenMX and GGA in DMol$^3$ can describe the energy diagrams of Ir$_2$ and Ir$_2$X correctly, considering the constraint of the special symmetries in these clusters. Both LDA and GGA in VASP cannot produce correct energy diagrams (see Fig. S3), which implies that the approaches based on plane-waves are not suitable for studying very small clusters like Ir$_2$ and Ir$_2$X. On the other hand, DMol$^3$ lacks the capability to calculate MAE. Hence we adopted the LDA implemented in OpenMX to calculate the structural, electronic and magnetic properties of Ir$_2$ and Ir$_2$X. As discussed in the main text, our present results on Ir$_2$ from LDA calculation with OpenMX agree well with previous experimental and theoretical data.

**References**
[1]  P. Hohenberg and W. Kohn, Physical review **136**, B864 (1964).
[2]  W. Kohn and L. J. Sham, Physical review **140**, A1133 (1965).
[3]  G. Bachelet, D. Hamann, and M. Schlüter, Phy. Rev. B **26**, 4199 (1982).
[4]  N. Troullier and J. L. Martins, Phy. Rev. B **43**, 1993 (1991).
[5]  L. Kleinman and D. Bylander, Phys. Rev. Lett. **48**, 1425 (1982).
[6]  P. E. Blöchl, Phy. Rev. B **41**, 5414 (1990).
[7]  I. Morrison, D. Bylander, and L. Kleinman, Phy. Rev. B **47**, 6728 (1993).
[8]  T. Ozaki, Phy. Rev. B **67**, 155108 (2003).
[9]  T. Ozaki and H. Kino, Phy. Rev. B **69**, 195113 (2004).
[10] J. P. Perdew, K. Burke, and M. Ernzerhof, Phy. Rev. Lett. **77**, 3865 (1996).
[11] G. Kresse and J. Furthmüller, Phy. Rev. B **54**, 11169 (1996).
[12] G. Kresse and D. Joubert, Phy. Rev. B **59**, 1758 (1999).
[13] P. E. Blöchl, Phy. Rev. B **50**, 17953 (1994).
[14] B. Delley, J. Chem. Phys. **92**, 508 (1990).
[15] B. Delley, J. Chem. Phys. **113**, 7756 (2000).